\documentclass[showpacs,aps,pra,twocolumn]{revtex4}
\usepackage{graphicx,epsfig}
\usepackage{times}
\usepackage{amsmath,amssymb}

\newlength\figurewidth
\newcommand{\rem}[1]{}

\newcommand{\imag}[1]{\text{Im}(#1)}

\newcommand{\real}[1]{\text{Re}(#1)}

\newcommand{\neff}{n}

\begin{document}

\title{Formation of long-lived, scarlike modes near avoided resonance
  crossings in optical microcavities}
  \author{Jan Wiersig}
  \affiliation{Institut f{\"u}r Theoretische Physik, Universit{\"a}t Bremen,
  Postfach 330 440, D-28334 Bremen, Germany}
\date{\today}

\begin{abstract}
We study the formation of long-lived states near avoided resonance crossings in
open systems. For three different optical microcavities (rectangle, ellipse,
and semi-stadium) we provide numerical evidence that these
states are localized along periodic rays, resembling scarred states
in closed systems. Our results shed light on the morphology
of long-lived states in open mesoscopic systems. 

\end{abstract}
\maketitle

Understanding the properties of long-lived, quasi-bound states in open
mesoscopic systems is of central importance for many research subjects, 
e.g. semiconductor ballistic quantum dots~\cite{AFB96,AFB97,ZB97,FAB04},
photoionization of Rydberg atoms~\cite{JSSS89}, microwave
systems~\cite{HHH99,Stoeckmann00}, quantum chaos~\cite{BMHK02}, and optical
microcavities~\cite{ND97,GCNNSFSC98,LLCMKA02,RTSCS02,HCLL02,LRRKCK04,FYC05}. 
In several of these studies the long-lived states are {\it scarred}. The 
original scarring phenomenon has been discovered for closed systems in the
field of quantum chaos~\cite{Heller84}. It refers to the existence of
a small fraction of eigenstates with strong localization along unstable
periodic orbits of the underlying classical system. In open systems, however,
scarred states seem 
to be the rule rather than the exception. The nature of the mechanism behind
this scarlike phenomenon is not yet understood.

Avoided level crossings in closed or conservative systems are discussed in
textbooks on 
quantum mechanics. They occur when the curves of two energy eigenvalues, as
function of a real parameter $\Delta$, come near to crossing but then
repel each other~\cite{NW29}.  
This behaviour can be understood in terms of a $2\times 2$
Hamiltonian matrix
\begin{equation}\label{eq:twolevelmodel}
H = \left(\begin{array}{cc} E_1 & V \\ W & E_2
\end{array}\right) \ .
\end{equation}
For a closed system this matrix is Hermitian, thus the energies $E_j$ are real
and the complex off-diagonal elements are related by $W = V^*$. The
eigenvalues of the coupled system,  
\begin{equation}\label{eq:eigenvalues}
E_\pm(\Delta)  = \frac{E_1+E_2}{2}\pm\sqrt{\frac{(E_1-E_2)^2}{4}+VW} \ ,
\end{equation}
differ from the energies of the uncoupled system $E_j$ only in a narrow
parameter region where the detuning from resonance,
$|E_1(\Delta)-E_2(\Delta)|$, is smaller or of the size of the coupling
strength $\sqrt{VW}$. The parameter dependence of $V$ and $W$ can often be
safely ignored.  

The matrix~(\ref{eq:twolevelmodel}) also captures features of
avoided {\it resonance} crossings (ARCs) in open or dissipative systems if one allows for complex-valued
energies $E_j$. The imaginary part determines  the lifetime $\tau_j \propto
1/\imag{E_j}$ of the quasi-bound state far away from the avoided crossing
$|E_1-E_2|^2 \gg VW$, where the off-diagonal coupling can be neglected. Keeping 
the restriction $W = V^*$ allows for two different kinds of
ARCs~\cite{Heiss00}.  
For $2|V| > |\imag{E_1}-\imag{E_2}|$, there is an avoided crossing in the real
part of the energy and a crossing in the imaginary part. At resonance
$\real{E_1} = \real{E_2}$ the eigenvectors of the
matrix~(\ref{eq:twolevelmodel}) are symmetric and antisymmetric
superpositions of the eigenvectors of the uncoupled system. If one of the latter
corresponds to a localized state then such an ARC leads to
delocalization and lifetime shortening~\cite{TR99}. 
For $2|V| < |\imag{E_1}-\imag{E_2}|$, there is a crossing in the real part
and an avoided crossing in the imaginary part. This
kind of ARC has been exploited to design optical microcavities with
unidirectional light emission from long-lived states~\cite{WH06}. 

\begin{figure}[t]
\includegraphics[width=0.95\figurewidth]{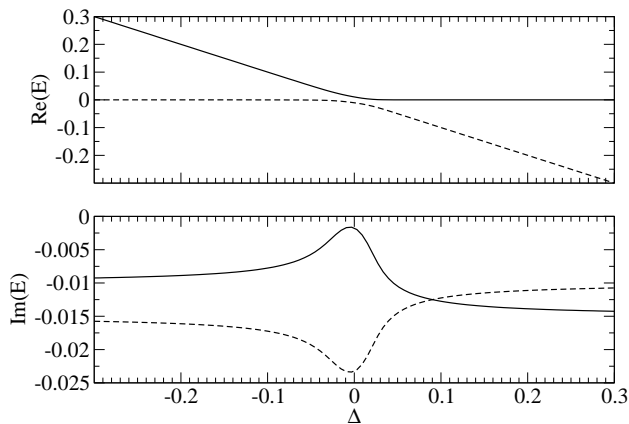}
\caption{Avoided resonance crossing in the case of the matrix~(\ref{eq:twolevelmodel}) with $VW =
  i0.000225$, $E_1 = -i0.015$, and $E_2 =
  \Delta-i0.01$. Real (top) and imaginary (bottom) part 
  of the energy vs. parameter $\Delta$.}   
\label{fig:arc2levelsystem}
\end{figure}

The case $W = V^*$ is called {\it internal coupling} since the only
difference to the Hermitian coupling of two states in a closed system is that
each state is {\it individually} coupled to the continuum. The 
latter is described by the imaginary part of the diagonal elements of
matrix~(\ref{eq:twolevelmodel}). The fully nonhermitian case $W\neq V^*$ is
more general; it permits an {\it external coupling} of the states  
{\it via} the continuum. Figure~\ref{fig:arc2levelsystem}
illustrates that the real part undergoes an avoided level crossing as in the
case of a closed system. The important feature is that one 
of the states has a considerably increased lifetime.
The constraint of the conservation of the trace of the matrix in
Eq.~(\ref{eq:twolevelmodel}) simultaneously
generates a state with short lifetime. The formation of fast and slowly
decaying states is known as resonance trapping, see, e.g.,
Refs.~\cite{DJ95,PRSB00}. 

The aim of this letter is to show that ARCs due to external coupling can have
a strong impact on the localization properties of long-lived states in open 
systems. The symmetric or antisymmetric superpositions are more localized in
real or phase space than the original states, so that important decay
channels are blocked. 
A surprising finding is that these states can resemble scarred states which
helps to explain the frequently observed scarring in open mesoscopic systems.

We examine optical microcavities, where the optical modes and their
frequencies play the role of states and their energies. Light confinement in
microcavities has attracted considerable interest in recent years due
to the huge potential for various research fields and applications, for a
review see Ref.~\cite{Vahala03}. For most applications, like low-threshold
lasing, long-lived modes are required. 
We consider quasi-two-dimensional dielectric cavities
with rectangular, elliptical and stadium-shaped cross section. We first focus
on rectangles because of the following convenient properties:  
(i) the modes not close to an ARC can be computed analytically to a good
approximation, including mode pattern and complex frequency;   
(ii) the internal ray dynamics is trivial, so localization effects
related to chaotic ray dynamics~\cite{LRKRK05} can be ruled out.
Rectangular and square microcavities have already been
studied both experimentally and theoretically~\cite{PCC01,Lohmeyer02,GHLY03}.
However, ARCs in these systems have not been addressed so far.
%

We fix one side length to $R=2\mu$m and vary the aspect ratio $0<\varepsilon
\leq 1$. We choose the  
effective index of refraction to be $\neff = 3.3$ inside and $\neff = 1$ 
outside the dielectric for the transverse electric (TE) polarization with
magnetic field $\vec{H}$ perpendicular to the cavity plane. Maxwell's
equations for the modes $H_z(x,y,t) = \psi(x,y)e^{-i\omega t}$ reduce to
a two-dimensional  scalar wave equation~\cite{Jackson83}
\begin{equation}\label{eq:wave}
-\nabla^2\psi = \neff^2(x,y)\frac{\omega^2}{c^2}\psi \ ,
\end{equation}
with frequency $\omega$ and the speed of light in vacuum $c$. 
The wave function $\psi$ and its normal derivative times $\neff^{-2}$
are continuous across the boundary of the cavity. At infinity, 
outgoing wave conditions are imposed. 
Even though the geometry of the cavity is rather simple the wave equation
cannot be solved analytically since the boundary conditions introduce
diffraction at corners. We compute the modes numerically using the boundary
element method~\cite{Wiersig02b}. In order to apply this method each corner
is replaced by a quarter of a circle with radius much smaller than the
wavelength. We have carefully checked that the rounding does
not influence the solutions in the studied frequency 
regime. 

\begin{figure}[t]
\includegraphics[width=0.95\figurewidth]{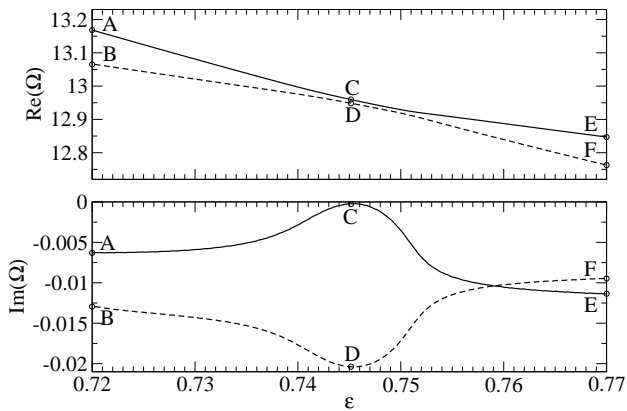}
\caption{An avoided resonance crossing in the rectangular microcavity. Plotted
 are the complex frequencies $\Omega$ as function of the aspect ratio $\varepsilon$.} 
\label{fig:arc}
\end{figure}

Before we discuss the numerical solutions of the open cavity we
will briefly consider the corresponding closed cavity with vanishing wave
intensity along the boundary. We expect that the solutions of the closed
system approximate those modes in the open system which are confined by total
internal reflection. The closed cavity is called an integrable
billiard~\cite{Stoeckmann00} since the modes can be computed analytically   
\begin{equation}\label{eq:billiard}
\psi_{n_x,n_y}(x,y) = \sin{\left(\frac{\pi
    n_x}{R}x\right)}\sin{\left(\frac{\pi n_y}{\varepsilon R}y\right)} 
\end{equation}
if $0\leq x \leq R$ and $0\leq y \leq \varepsilon R$; otherwise
$\psi_{n_x,n_y}(x,y)=0$. The positive integers $n_x$, $n_y$ count the number
of maxima of 
$|\psi|^2$ in $x$- and $y$-direction, respectively. The normalized frequency
$\Omega = \omega R/c$ belonging to such a mode is given by 
\begin{equation}\label{eq:omega}
\Omega_{n_x,n_y} = \frac{\pi}{\neff}\sqrt{n_x^2+\frac{n_y^2}{\varepsilon^2}} \ .
\end{equation}
As expected for an integrable billiard~\cite{Stoeckmann00}, this system shows
frequency crossings instead of avoided crossings when the aspect ratio
$\varepsilon$ is varied. For example, for the modes $(n_x,n_y) = (10,7)$ and
$(12,5)$ Eq.~(\ref{eq:omega}) yields the crossing point
$\varepsilon = \sqrt{6/11} \approx 0.739$ and $\Omega \approx 13.12$
corresponding to a free-space wavelength of about 960nm. 
Figure~\ref{fig:arc} shows that this accidental degeneracy is lifted in the
open cavity. The associated ARC equals the case of the $2\times 2$ matrix in
Fig.~\ref{fig:arc2levelsystem}. We therefore conclude that diffraction at
corners in rectangular cavities leads to an external coupling of modes.

\begin{figure}[t]
\includegraphics[width=0.95\figurewidth]{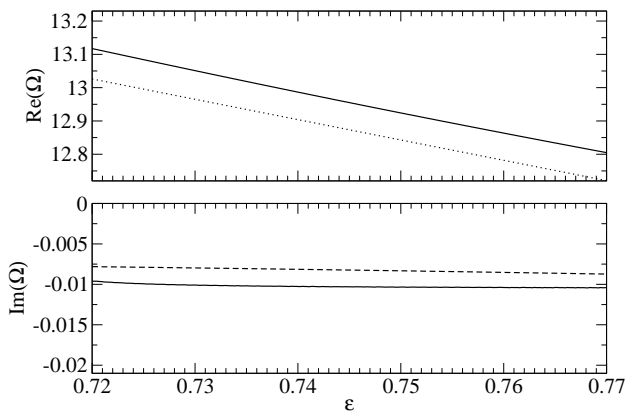}
\caption{Mean value of real (top) and imaginary (bottom) part of frequencies
(solid lines) near the avoided resonance crossing shown in
  Fig.~\ref{fig:arc}. The dotted line is the scar
  approximation~(\ref{eq:scar}). The dashed line is the averaged result of
  the boundary 
  wave approach~(\ref{eq:bw}). } 
\label{fig:arcmean}
\end{figure}

It has been demonstrated in Ref.~\cite{Wiersig03} that losses from a polygonal
cavity due to diffraction at corners can be estimated by the boundary wave
approach. Boundary 
waves travel along a flat interface between dielectric material and air. In the
case of an infinitely extended interface, these waves are evanescent. In the
case of a finite interface, however, these waves can leave the interface
region at the corners. Following Ref.~\cite{Wiersig03} we have derived a
formula describing the losses from a given mode $(n_x,n_y)$ in the rectangular
cavity due to boundary waves  
\begin{eqnarray} \label{eq:bw}
\imag{\Omega} = -\frac{2\neff}{\varepsilon\;\real{\Omega}}\sum_{j=1}^2\frac{\sin{\theta_j}}{\sqrt{\neff^2\sin{\theta_j}^2-1}\;(1+\alpha_j^2)}
\end{eqnarray}
with $\alpha_j =
{\neff\sqrt{\neff^2\sin{\theta_j}^2-1}}/{\cos{\theta_j}}$, 
$\tan{\theta_1} = \varepsilon n_x/n_y$, and $\theta_2=\pi/2-\theta_1$.
For a mode A with $\varepsilon = 0.72$ and $(n_x,n_y) = (10,7)$ we find 
$\imag{\Omega} \approx -0.0048$ corresponding to a quality factor of
$Q=\real{\Omega}/[2\imag{\Omega}] \approx 1380$. For a mode B with
$\varepsilon = 0.72$ and $(n_x,n_y) = (12,5)$ we get $\imag{\Omega} 
\approx -0.0109$ and $Q\approx 600$. 
The boundary wave approach as developed in
Ref.~\cite{Wiersig03} can only compute the losses of
individual modes, i.e. the diagonal elements of the
matrix~(\ref{eq:twolevelmodel}). The off-diagonal part, i.e. the coupling of
modes, cannot be determined
within this approach. For a direct comparison to the exact results in
Fig.~\ref{fig:arc} it is therefore useful to  consider the mean value of
$\Omega_+$ and $\Omega_-$ since here the ARC contributions cancel; cf. 
Eq.~(\ref{eq:eigenvalues}). The result of this procedure is shown in the bottom
panel of Fig.~\ref{fig:arcmean}. It can be seen that the averaged boundary 
wave result overestimates the averaged lifetimes of the modes by just 20
percent. Hence, leakage due to boundary waves is the dominant decay
channel. 

At the center of the ARC, $\varepsilon \approx 0.7453$, in Fig.~\ref{fig:arc}
a fast mode D with $\imag{\Omega} \approx -0.02$ and a slow mode C with
$\imag{\Omega} \approx -0.00028$ is formed. The slow mode has $Q\approx
23\,200$ 
which is a dramatic increase by more than one order of 
magnitude if compared to the ``normal'' quality factor.   
In this frequency regime the leakage due boundary waves limits the quality
factor to roughly $1900$. This indicates that possibly all long-lived modes
(modes with, say, $Q\geq 4000$) in this frequency regime are caused
by ARCs. This conclusion is supported by extensive numerical studies on this
system (not shown). 

\begin{figure}[t]
\includegraphics[width=1.0\figurewidth]{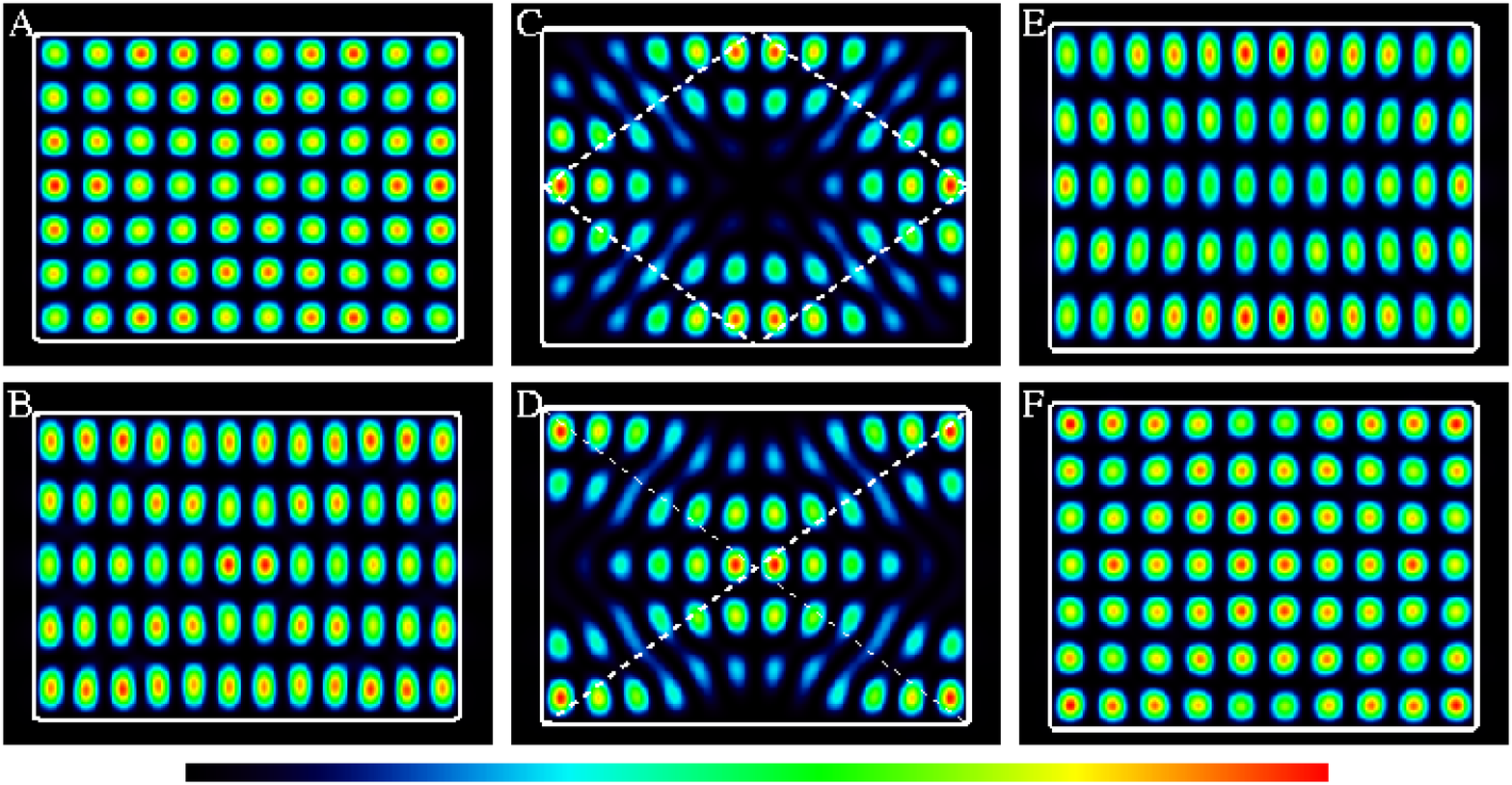}
\caption{(Color online) Calculated near field intensity of modes with the same
  labels as in Fig.~\ref{fig:arc}. Mode C shows localization along a periodic
  ray (dashed line), mode D is localized along two rays connecting the corners
  (dashed lines).}
\label{fig:modes}
\end{figure}

The spatial patterns of modes A, B, E, and F in Fig.~\ref{fig:modes} 
approximately match the solutions of the closed cavity in 
Eq.~(\ref{eq:billiard}). Upon the avoided crossing the mode
patterns exchange their character, i.e. mode B and E have roughly the same
spatial profile but belong to different frequency branches;
cf. Fig.~\ref{fig:arc}. The same holds for mode A and F.  
The modes at the ARC, C and D, correspond to symmetric and antisymmetric
superpositions of the mode A and B (or E and F). Now, we can identify the
physical 
mechanism behind the increased quality factor of mode C: destructive
interference reduces the light intensity at the corners and consequently
the  main decay channel, leakage due to boundary waves, is strongly
suppressed.    
A closer inspection of the mode C in Fig.~\ref{fig:modes} reveals that
its intensity is concentrated along a diamond-shaped periodic ray. The
long-lived mode formed in the ARC therefore resembles a scarred mode.
In the case of mode D, constructive interference at the corners spoils the 
quality factor. 
Mode D is localized along two symmetry-related rays connecting the corners of
the cavity. Such kind of rays are called diffractive rays~\cite{HHH99}. 

The spatial pattern of mode C is not a special feature of the chosen boundary
conditions but is also observed in square cavities of very different types: 
(i) square quantum dots with leads attached at the corner region~\cite{AFB96};
(ii)  square billiards with magnetic flux~\cite{NPZ00}; 
(iii)  vertical-cavity surface-emitting lasers with square-shaped cross
section~\cite{HCLL02,CHLL03}.   

The resemblance of mode C with the diamond-shaped 
periodic ray is apparent with the naked eye. In the following, it will be
demonstrated that the relation is even deeper. To do so, we estimate the 
frequency of the mode by using the localization along the ray. 
We stipulate that an integer number $m$
of wavelengths fits onto the periodic ray with length 
$l=2R\sqrt{1+\varepsilon^2}$. The calculation is straightforward giving
\begin{equation}\label{eq:scar}
\Omega_{\mbox{\footnotesize scar}} = \frac{\pi}{\neff\sqrt{1+\varepsilon^2}}(m+\beta)
\end{equation}
with $\beta = \frac{2}{\pi}\sum_{j}\arctan{\alpha_j}$ being the total phase
shift from the reflection at the dielectric boundary for TE polarization.
The quantities $\alpha_1$ and $\alpha_2$ are the same as for Eq.~(\ref{eq:bw})
but with $\tan{\theta_1} = 1/\varepsilon$,  
and $\theta_2 = \pi/2-\theta_1$. The top panel of Fig.~\ref{fig:arcmean}
demonstrates that the scar approximation with $m=15$ describes the mean
behaviour of the modes involved in the ARC over a broad range of parameter
values.  
The small frequency offset of about $0.086$ can be traced back to the
fact that the scar approximation assumes that the plane waves lying on
the ray segments have no wave vector component in the transverse
direction. However, the mode is restricted to an interval of 
length $R$ ($\varepsilon R$) in $x$-direction ($y$-direction). At least half a
wavelength fits into these intervals for which the wave vector components 
give frequency contributions of
approximately $\pi/\neff$ and $\pi/(\varepsilon \neff)$. Summing up the
squares of the frequency contributions gives the correction
$\Delta\Omega \approx
{\pi^2\left(1+{1}/{\varepsilon^2}\right)}/{2\neff^2\Omega_{\mbox{\footnotesize
      scar}}}$.
In the regime $\varepsilon = [0.72,0.77]$ we get $\Delta\Omega \approx
0.096..0.102$ which convincingly explains the discrepancy between
$\Omega_{\mbox{\footnotesize scar}}$ and the exact value of~$\Omega$. 

Coupling between two modes $(n_x,n_y$) and $(m_x,m_y)$ occurs only
for modes with the same symmetry with respect to the lines $x
= R/2$ and $y=\varepsilon R/2$. That implies that if $n_x$ is even (odd) $m_x$
must be even (odd) too. The same holds for $n_y$ and $m_y$. Interestingly,
this restriction ensures that for given allowed pair $(n_x,n_y$) and
$(m_x,m_y)$ cancellation at {\it all} corners is possible.

\begin{figure}[t]
\vspace{0.2cm}
\includegraphics[width=1.0\figurewidth]{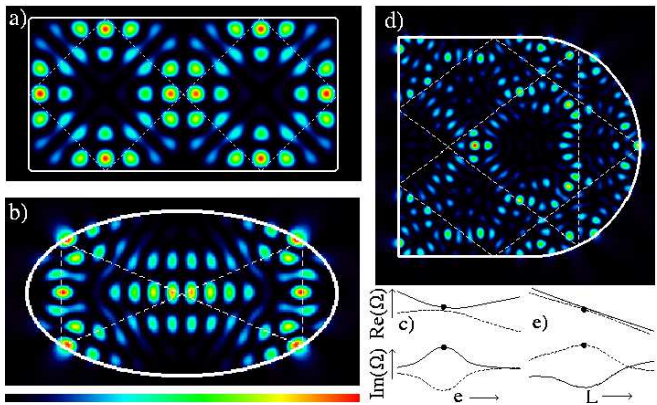}
\caption{(Color online) a) Near field intensity of a long-lived mode resulting
  from an ARC of modes $(10,7)$ and $(14,5)$.  The white dashed line is the
  periodic ray. b) and c) Long-lived TM mode
  formed at an ARC (marked by the dot) in an elliptical
  resonator with eccentricity~$e$. d) and e) Long-lived TM mode at an ARC in a
  stadium resonator with horizontal lines of length~$L$.}  
\label{fig:higherorderscar}
\end{figure}

We can create a variety of scarlike modes near ARCs. Consider a periodic ray
bouncing $q$ times at the horizontal lines and $p$ times at the vertical
lines. A straightforward analysis shows that $q = 2[(|n_x-m_x|+1)/2]$ and $p =
2[(|n_y-m_y|+1)/2]$ where $[\ldots]$ denotes the integer part. 
Figure~\ref{fig:higherorderscar}(a) depicts an example with $q=4$ and $p=2$. 
This long-lived mode with $\Omega \approx 16.306-i0.00047$ ($Q \approx
17500$) results from an ARC of modes $(10,7)$ and $(14,5)$ at 
$\varepsilon = 0.4954$. 

The formation of long-lived, scarlike modes near ARCs with external coupling
is not restricted to TE polarization nor to the rectangular geometry. 
Figures~\ref{fig:higherorderscar}(b)-(e) show examples with transverse 
magnetic (TM) polarization in an elliptical and a (semi-) stadium-shaped
resonator with refractive index $\neff=3.3$. The shape parameter 
is the eccentricity $e$ and the length of the horizontal straight lines $L$,
respectively.  
The scenario is as in the cases shown in
Figs.~\ref{fig:arc2levelsystem} and \ref{fig:arc}. At the ARC a short- and a
long-lived mode is formed; see Fig.~\ref{fig:higherorderscar}(c) and (e). The
long-lived mode exhibits a localization along a periodic  
ray; see Fig.~\ref{fig:higherorderscar}(b) and (d). This localization
gradually disappears when the shape parameter is detuned from
resonance (not shown). 
Note that elliptical billiards do not show such
scarring~\cite{WWD97}.  
Let us mention that local maxima of quality factors as function of a shape 
parameter had already been exploited for minimizing losses from stadium-shaped
cavities~\cite{FYC05}. However, the case in Ref.~\cite{FYC05} is not related
to ARCs, but is an interference effect of unstable periodic
rays~\cite{FHW06}. 
  
In summary, we demonstrated the formation of long-lived modes near avoided
resonances crossings in optical microcavities. For a number of different types
of cavities (rectangular, elliptical, and stadium-shaped) we observed strong
localization of these modes, resembling scarred states.
We expect that this finding is highly relevant for understanding the
localization 
properties of long-lived states not only in optical systems but in various
fields of research. 

We acknowledge helpful discussions with M.~Hentschel, F.~Anders T.~Y.~Kwon, and
T.~Gorin.   

\bibliography{}

\end{document}